\documentclass[hyper]{JHEP} 

\usepackage{epsfig}

\newcommand\fverb{\setbox\pippobox=\hbox\bgroup\verb}

\newcommand\fverbdo{\egroup\medskip\noindent%

            \fbox{\unhbox\pippobox}\ }

\newcommand\fverbit{\egroup\item[\fbox{\unhbox\pippobox}]}

\newbox\pippobox

\title{Algebra of Lax Connection
for T-Dual Models}

\author{J. Kluso\v{n}\\
Department of
Theoretical Physics and Astrophysics\\
Faculty of Science, Masaryk University\\
Kotl\'{a}\v{r}sk\'{a} 2, 611 37, Brno\\
Czech Republic\\
E-mail: \email{klu@physics.muni.cz}}
\preprint{ \hepth{0812.4510}}
\abstract{We study
  relation between
T-duality and integrability. We develop
the Hamiltonian formalism for principal
chiral model on general group manifold
and on its T-dual image. We calculate
the Poisson bracket of Lax connections
in T-dual model and we show that they
are non-local as opposite to the
Poisson brackets of Lax connection in
original model. We demonstrate these
calculations on two specific examples:
Sigma model on $S^2$ and sigma model on
$AdS_2$.}


\newcommand{\mT}{\mathcal{T}}
\def\tr{\mathrm{Tr}}
\def\tJ{\tilde{J}}

\newcommand{\tX}{\tilde{X}}

\def\pb  #1{\left\{#1\right\}}

\def\hr{\hat{r}}

\newcommand{\bA}{{\bf A}}

\newcommand{\bC}{\mathbf{C}}

\newcommand{\mG}{\mathcal{G}}
\newcommand{\mA}{\mathcal{A}}

\newcommand{\mE}{\mathcal{E}}
\newcommand{\mF}{\mathcal{F}}
\newcommand{\teta}{\tilde{\eta}}

\newcommand{\hJ}{\hat{J}}

\newcommand{\tOmega}{\tilde{\Omega}}

\newcommand{\bJ}{\mathbf{J}}

\newcommand{\tPhi}{\tilde{\Phi}}

\newcommand{\bB}{\mathbf{B}}
\newcommand{\tbA}{\tilde{\bA}}
\newcommand{\tbB}{\tilde{\bB}}
\newcommand{\tbC}{\tilde{\bC}}
\newcommand{\tY}{\tilde{Y}}
\begin{document}
\section{Introduction and Summary}
One of the most remarkable achievements
in string theory that were reached in
past few years is a discovery of
integrability of $N=4$ superconformal
Yang-Mills theory in its planar limit
together with  integrability of
$AdS_5\times S^5$ superstring
\cite{Beisert:2004ry,Belitsky:2004cz,Minahan:2002ve}
\footnote{For review and extensive list
of references, see
\cite{Vicedo:2008jk,Minahan:2006sk,Okamura:2008jm,
Tseytlin:2003ii,Plefka:2005bk,Zarembo:2004hp}.}.
Indeed, the well-known classical
integrability of bosonic string on
$AdS_5\times S^5$ was extended to
$\kappa$-symmetric Green-Schwarz
superstring
\cite{Bena:2003wd,Alday:2005gi}, or to
the pure spinor formulation of
superstring as well
\cite{Magro:2008dv,Puletti:2008ym,Berkovits:2008qc,
Linch:2008nt,Mikhailov:2007mr,Mikhailov:2007eg,
Berkovits:2007zk,Grassi:2006tj,Puletti:2006vb,Bianchi:2006im,
Chandia:2006ix,Berkovits:2004xu,Berkovits:2004jw,Vallilo:2003nx,
Vallilo:2002mh} \footnote{For review,
see
\cite{Bedoya:2008yw,Guttenberg:2008ic,Berkovits:2002zk,Grassi:2003cm}}.
For example, it was found that
the classical superstring
possesses an  infinite
number  of conserved non-local charges.
 These
charges  have their  counterpart in
planar gauge theory at weak coupling in
the spin-chain formulation for the
dilatation operator
\cite{Dolan:2004ps,Dolan:2003uh}.

It is also  believed that the integrability
of the $N=4$ SYM should have an
impact on the spectrum of other
observables in the theory, for example
on the structure of expectation values
of certain Wilson loops.  The dual
formulation of these objects in
$AdS/CFT$ correspondence are partition
functions of \emph{open} $AdS_5\times
S^5$ strings that end on some contours
at the boundary of $AdS_5$
\cite{Maldacena:1998im,Rey:1998bq}.
Moreover, it turned out that the open
string description of Wilson loops is
directly related to the T-duality in
$AdS_5\times S^5$
\cite{Alday:2007hr,Alday:2007he}
 \footnote{For review and extensive list of
 references, see
\cite{Alday:2008yw,Alday:2008cg,Makeenko:2008xr}.},
where T-dual formulation appears to be
important in discovery of connection
between maximally helicity-violating
(MHV) gluon scattering amplitudes and
special Wilson loops (defined on
contours formed by light-like gluon
momentum vectors). The classical
$SO(4,2)$ conformal symmetry of the
T-dual $AdS_5$ geometry seems to be
related to a mysterious "dual"
conformal symmetry that was observed in
the momentum-space integrands of loop
integrals for planar gluon scattering
amplitudes
\cite{Bern:2006ew,Drummond:2008vq,
Brandhuber:2007yx,Drummond:2007aua}.
From the string theory point of view
this dual conformal symmetry of
$AdS_5\times S^5$ sigma model could be
related to the presence of hidden
symmetries in T-dual string theory.
In fact, since T-duality is an
on-shell symmetry, or in other words it
maps classical solutions to classical
one and since the T-dual geometry is
again $AdS_5\times S^5$ we can expect
that T-dual model is also integrable
and consequently possesses an infinite
number of conserved charges that
should correspond to
generators of some  symmetries
of dual  Wilson lines in $N=4$ SYM.

In phase space formulation
of string theory the
statement that  T-duality is on-shell
symmetry is that
T-duality is  canonical transformation
\cite{Lozano:1996sc,Alvarez:1994wj}.
This fact has a significant
consequence for the calculation of the
Poisson brackets of Lax connection in
T-dual theory. In fact, it is
 an important
question  how the Lax connection in
original integrable theory  is mapped
to its T-dual counterpart \footnote{For
some earlier works that discuss this
problems, see
\cite{Frolov:2005dj,Hatsuda:2006ts,Kluson:2007md,
Gomes:2004bw,Miramontes:2004dr,
Balog:1993es,Balog:1996zx,Balog:2000wk,Forgacs:2000eu}.}.
 These problems were discussed recently in
very nice papers \cite{Ricci:2007eq,
Beisert:2008iq,Berkovits:2008ic}. In
particular, the paper
\cite{Ricci:2007eq} discussed the
interplay between T-duality and
integrability on two examples: the
two-sphere $S^2$ and the
two-dimensional anti-de Sitter space
$AdS_2$. It was argued there that in
order to perform T-duality explicitly
we have to express  T-dual
coordinates in terms of the original
ones. Further, it is also
 important that this
relation is non-local and hence it is a
non-trivial task to find the T-duality
image of Lax connection since Lax
connection-as opposite to the sigma
model action-depends explicitly on
coordinate that parameterizes T-duality
direction. On the other hand it was
shown in \cite{Ricci:2007eq} that it is
possible to eliminate the explicit
dependence of Lax connection on the
T-duality direction coordinate with the
help of some special field redefinition
that preserves the flatness of Lax
connection of original theory. The new
Lax connection depends on the
derivatives of the isometric coordinate
only. Then  T-duality  on these flat
currents can be easily implemented and
then it is easy to find the T-dual flat
currents.

The goal of this paper is to study
properties of the Lax connection in
T-dual background further. We
are mainly interested in the calculation of
the Poisson bracket of the T-dual Lax
connections. We find explicit form of
these Poisson brackets and we argue
that they have exactly the same form as
the brackets introduced in
\cite{Maillet:1985ek,Maillet:1985ec}.
Then, following discussion presented in
these papers and reviewed in appendix
 we can argue that
 T-dual  theory possesses an infinite
number of
conserved charges that are  in
involution \footnote{Of course, up the
subtlety of the definition of the
Poisson bracket of monodromy matrices
in case when the initial and final
points coincide. For discussion of
these problems, see
\cite{Maillet:1985ek,Maillet:1985ec,Dorey:2006mx,
Duncan:1989vg}.} in the sense that
their Poisson brackets commute.
 On the other hand it
is desirable to find an explicit form
of matrices $r,s$ that appear in these
Poisson brackets and that are crucial
for the study of classical or quantum
mechanical integrability of given
theory
\cite{Maillet:1985ek,Maillet:1985ec}.
We  show that even if it is
rather straightforward to find the form
of these matrices in case of principal
chiral model  it is very
difficult to find their  form  in
T-dual model. In fact, we will argue
that the original constant matrices
$r,s$ map under T-duality to non-local
expressions. We claim that this is in
accord with the observation that local
charges map to non-local ones under
T-duality.

 The organization of this
paper is as follows. In the next
section (\ref{second}) we review the
calculation of Poisson bracket of Lax
connection  in case of principal chiral
model and we show that in this case the
 matrices $r,s$ are constant. Then
in section (\ref{third}) we perform
calculation of Poisson bracket of Lax
connection for the sigma model that is
T-dual to the sigma model on $S^2$. We
review the construction of Lax
connection  performed  in
\cite{Ricci:2007eq} and express it as
function of canonical variables. Then
we calculate the Poisson brackets of
Lax connection for T-dual action
and argue for integrability of
given theory. In
section (\ref{fourth}) we study another
example of T-dual sigma model that
arises by performing T-duality along
non-compact direction of $AdS_2$.
 We again calculate  the Poisson
brackets of Lax connection that is
an image of  original Lax connection
under T-duality.
 We stress however that  T-duality in
case of $AdS_2$ background is special
since now T-duality along non-compact
direction leads  to background
that it is again $AdS_2$. Then it is natural
to presume that this background
possesses Lax connection that has the
same form as the original one now
expressed as  a function of T-dual
variables. One can then expect
that these two Lax connections are
related and it was shown for example
 in  \cite{Beisert:2008iq} this
is really true.  We plan to
discuss the Hamiltonian formulation of
these Lax connections and relations
between them in forthcoming
publication. Finally in section
(\ref{gener}) we present the general
analysis of the calculation of Poisson
bracket between Lax connections in
sigma model that is related to the original
one by duality that is
canonical transformation in phase space.
We define the Lax connection in dual
model in two steps. We firstly perform
the gauge transformation that preserves
the flatness of given Lax connection. In
the second step we express this Lax
connection as a functions of phase space
variables that are related to the original
ones by canonical transformations.
 In
fact, it is well known that T-duality
can be considered as some kind of
canonical transformations. We
 find that the
Poisson brackets of dual Lax connections
 are non-local however their forms
 again imply that the new theory
possesses an infinite number of conserved
charges that are in involution.

Let us outline our result. We derive
Poisson brackets of Lax connections in
sigma models that are T-dual to original
integrable models. We argue that the
new integrable models contain an infinite
number of conserved charges that are in
involutions even if the Poisson brackets
of Lax connections are non-local and
the matrices $r,s$ are functions of
phase space variables. We hope that these
calculations can be useful for further
study of relations between T-duality
and integrability.

\section{Review of Principal Chiral Model
$S^2$ and Poisson Bracket of its Lax
Connection}\label{second}
In this section we give a brief   review
of calculation of Poisson bracket of Lax
connection of the sigma model
 on $S^2$. The main goal
is to demonstrate the difference
between rather straightforward
calculations given here with respect to
the analysis of the Poisson bracket of
Lax connection in case of T-dual model
that will be presented in next section.

We start with the action that governs
the dynamics
 of the string on $S^2$
\begin{equation}\label{S2act}
S=-\frac{1}{2} \int d^2\sigma
\sqrt{-\gamma} \gamma^{\alpha\beta}
[\partial_\alpha \Theta\partial_\beta
\Theta+\sin^2\Theta \partial_\alpha
\Phi
\partial_\beta \Phi] \ ,
\end{equation}
where $\gamma_{\alpha\beta} \ ,
\alpha,\beta=0,1$ is world-sheet
metric and where $\sigma^\alpha \ ,
\sigma^0=\tau \ , \sigma^1=\sigma$
are world-sheet coordinates.

We start with observation
that  this theory possesses Noether
currents in the form
\begin{eqnarray}
j^1&=&-\frac{1}{\sqrt{2}} [\sin(\Phi)
d\Theta+ \sin \Theta \cos\Theta
\cos\Phi d\Phi] \ , \nonumber \\
j^2&=&-\frac{1}{\sqrt{2}} [\cos\Phi
d\Theta-\sin\Theta \cos\Theta \sin\Phi
d\Phi] \ , \nonumber \\
j^3&=&-\frac{1}{\sqrt{2}} \sin^2\Theta
d\Phi  \ , \nonumber \\
\end{eqnarray}
where $df\equiv \partial_\alpha f d\sigma^\alpha$.
Then it is easy to see that
\begin{eqnarray}\label{jc}
\gamma^{\alpha\beta}j^A_\alpha
j^B_\beta K_{AB}=
-\frac{1}{2}\gamma^{\alpha\beta}
[\partial_\alpha \Theta\partial_\beta
\Theta+ \sin^2\Theta \partial_\alpha
\Phi
\partial_\beta \Phi] \ ,
\nonumber \\
\end{eqnarray}
where $A,B,\dots=1,2,3$ and where the
Cartan-Killing form is given
by
\begin{equation} K_{AB}=\tr (T_A
T_B)=\mathrm{diag}(-1,-1,-1) \ .
\end{equation}
Then using this result we can
rewrite the  action  (\ref{S2act})
in the form
\begin{equation}
S=\int d^2\sigma \sqrt{-\gamma}
\gamma^{\alpha\beta}\tr j_\alpha j_\beta \
\end{equation}
that clearly demonstrates that the
dynamics of string on $S^2$ is governed
by principal model action. Further, we
observe that the currents (\ref{jc})
are flat
\begin{equation}
\partial_\alpha j^A_\beta-\partial_\beta
j_\alpha^A+j^B_\alpha j^C_\beta
f_{BC}^{ \quad A}=0 \ ,
\end{equation}
where the structure constants are defined
as
\begin{equation}
f_{BC}^{ \quad A}=
-\frac{1}{\sqrt{2}}\epsilon_{BCD}K^{DA} \ .
\end{equation}
Here $\epsilon_{ABC}$ is totally
antisymmetric with $\epsilon_{123}=-1$.
In what follows we will be more general
and introduce the general coordinates
$x^M$ on manifold $M$ that in the
particular case of  $M=S^2$ are
$x^M=(\Theta,\Phi)$. As the next step
we introduce  $E_M^A$ in order to write
the current $j^A$ in the form
\begin{equation}
j^A_\alpha= E^A_M\partial_\alpha x^M \ .
\end{equation}
Then the conjugate momenta $p_M$
defined as $\frac{\delta S}{\delta \partial_\tau
x^M}$ take the form
\begin{equation}
p_M=\sqrt{-\gamma}\gamma^{\tau\alpha}
K_{AB}E^A_M E^B_N\partial_\alpha x^N \ .
\end{equation}
Further  we define  the
 current $j^A_P$ as
\begin{equation}
j^A_P=-\sqrt{-\gamma}\gamma^{\tau\alpha}
E^A_M
\partial_\alpha x^M=-K^{AB}E_B^Mp_M \ ,
\end{equation}
where $E_B^M$ is inverse of $E^A_M$.
Then it is easy to calculate the
 Poisson bracket
\begin{eqnarray}\label{pbjSP}
\pb{j^A_\sigma(\sigma),j_P^B(\sigma')}&=&
-E^A_N(\sigma)
E^N_C(\sigma')K^{BC}\partial_\sigma
\delta(\sigma-\sigma')-\partial_NE^A_M
\partial_\sigma x^MK^{BC}E_C^A\delta(\sigma-\sigma')=
\nonumber \\
&=&-K^{AB}\partial_\sigma
\delta(\sigma-\sigma') -j^D_\sigma
f_{DC}^{ \quad
A}K^{CB}\delta(\sigma-\sigma') \
\nonumber \\
\end{eqnarray}
using
\begin{equation}
\partial_{\sigma'}\delta(\sigma-\sigma')=
-\partial_\sigma \delta(\sigma-\sigma') \ ,
\quad
f(\sigma')\partial_\sigma \delta(\sigma-\sigma')=
f(\sigma)\partial_\sigma \delta(\sigma-\sigma')+
\partial_\sigma f(\sigma)\delta(\sigma-\sigma') \
\end{equation}
and also the relations
\begin{equation}
\partial_N E_M^A-\partial_M E_N^A+
E^B_M E^C_N f_{BC}^{\quad A}=0 \ , \quad
\partial_P E^N_C=-E_C^M\partial_P E_M^DE_D^N \
\end{equation}
that follow from the fact
 that the current $j^A=E^A_M dx^M$ is flat.
In the same
way we obtain
\begin{eqnarray}\label{pbjPP}
\pb{j_P^A(\sigma),j_P^B(\sigma')}=
-j_P^C(\sigma) f_{CD}^{ \quad A}
K^{DB}\delta(\sigma-\sigma') \ .
\nonumber \\
\end{eqnarray}
Now we are ready to
 determine  the Poisson
brackets of Lax connection for
principal chiral model. Note that
 the Lax
connection is  defined as
\begin{equation}\label{flatorg}
J^A=aj^A+b*j^A \ ,
\end{equation}
where the Hodge dual is defined as
\begin{eqnarray}
(*df)_\alpha=
-\sqrt{-\gamma}\partial_\gamma f
\gamma^{\gamma\delta}\epsilon_{\delta\alpha} \
\end{eqnarray}
for any function $f$.
Further, $a,b$ given
in (\ref{pbjPP}) depend on
spectral parameter $\Lambda$ as
\begin{equation}
a=\frac{1}{2} [1\pm \cosh \Lambda] \ ,
\quad b=\frac{1}{2}\sinh \Lambda
\end{equation}
so that $a^2-a-b^2=0$.
  Explicitly, for spatial
components of $J_\sigma^A$ we obtain
\begin{eqnarray}\label{curorg}
J^A_{\sigma}= aj^A_{\sigma}-
b\sqrt{-\gamma}\gamma^{\tau\beta}j^A_{\beta}
=a j^A_\sigma+b j_P^A \ . \nonumber \\
\end{eqnarray}
Then using (\ref{pbjSP}) and
(\ref{pbjPP}) we  determine the Poisson
brackets between the spatial components
of Lax connections for two different
spectral parameters $\Lambda,\Lambda'$
\begin{eqnarray}\label{pbchir}
\pb{J^A_\sigma(\sigma,\Lambda),
J^B_\sigma(\sigma',\Lambda')}
&=&-ab'K^{AB}\partial_\sigma
\delta(\sigma-\sigma')+
ba'K^{AB}\partial_{\sigma'}\delta(\sigma-\sigma')-
\nonumber \\
&-& ab'j^D_\sigma f_{DC}^{ \quad A}
K^{CB}\delta(\sigma-\sigma') +ba'
j^D_\sigma f_{DC}^{ \quad B}
K^{CA}\delta(\sigma-\sigma')-
\nonumber \\
&-& bb'j_P^C f_{CD}^{ \quad A} K^{DB}\delta(\sigma-\sigma') \ . \nonumber \\
\end{eqnarray}
Comparing with the equation
(\ref{pbLaxa})  we find
\begin{eqnarray}
\bB^{AB}&=& ba' K^{AB} \ , \quad
\bC^{AB}=-ab' K^{AB} \ , \nonumber \\
\bA^{AB}&=& -ab' j^D_\sigma f_{DC}^{ \quad A}
K^{CB}-ba' j^D_\sigma f_{DC}^{ \quad A}
K^{CB}-bb'
j_P^C f_{CD}^{ \quad A} K^{DB} \ .  \nonumber \\
\end{eqnarray}
Fortunately in this particular  case we
can rather easily guess the form of the
matrices $r^{AB},s^{AB}$. In fact, let
us presume that the right side of the
equation (\ref{pbchir}) can be written
in the form
\begin{eqnarray}\label{guessform}
& &(r-s)^{DB}f_{DC}^{\quad A}
J^C_\sigma(\Lambda)+ (r+s)^{AD}f_{DC}^{\quad
B} J^C_\sigma(\Lambda')+
\nonumber \\
&+&\partial_\sigma
(r-s)^{AB}\delta(\sigma-\sigma')-2
s^{AB}
\partial_\sigma \delta(\sigma-\sigma')
\ .
\nonumber \\
\end{eqnarray}
Then comparing expressions proportional
to $\partial_\sigma
\delta(\sigma-\sigma')$ in
(\ref{pbchir}) and in (\ref{guessform})
we obtain
\begin{equation}\label{sAB}
s^{AB}=\frac{1}{2}K^{AB}(ab'+ba') \ .
\end{equation}
Further we presume that
\begin{equation}
(r-s)^{AB}=AK^{AB} \ , \quad
(r+s)^{AB}=BK^{AB} \ ,
\end{equation}
where $A,B$ are constants. Inserting
these expressions into (\ref{guessform})
and comparing with right side of
(\ref{pbchir}) we find
\begin{eqnarray}
B=-\frac{b^2a'}{a'b-b'a} \ , \quad
A=-\frac{b'^2 a}{a'b-b'a} \nonumber
\\
\end{eqnarray}
and then using (\ref{sAB}) we finally
obtain
\begin{equation}
r^{AB}=-\frac{1}{2(a'b-b'a)}
(b'^2a^2+a'^2b^2-2b'^2b^2) \ .
\end{equation}
It is important that in
case of principal chiral
model the objects  $r^{AB},s^{AB}$
are constants and depend on spectral
parameters $\Lambda,\Lambda'$ only.
 On the other hand
 the situation is much
more involved in case of T-dual sigma model.
\section{Poisson Brackets of Lax Connection
in T-dual theory on $S^2$}\label{third}
In this section we determine the
Poisson bracket of Lax connections
in the sigma model that is
related to the sigma model on $S^2$ by
T-duality along the compact $U(1)$
isometry cycle. Since the procedure how
to derive T-dual action from sigma
model action on $S^2$
 is nicely described in the paper
\cite{Ricci:2007eq}  we use results
derived there and immediately write the
T-dual action
\begin{eqnarray}\label{STdual}
S&=&-\frac{1}{2}\int (d\Theta *d\Theta
+\frac{1}{\sin^2\Theta}d\tPhi * d\tPhi)=
\nonumber \\
&=&-\frac{1}{2}\int d^2\sigma
\sqrt{-\gamma}
\gamma^{\alpha\beta}(\partial_\alpha
\Theta\partial_\beta \Theta+
\frac{1}{\sin^2\Theta}\partial_\alpha
\tPhi
\partial_\beta \tPhi) \ ,
\nonumber \\
\end{eqnarray}
where the dual variable $\tPhi$ is related
to $\Phi$ through the relation
\begin{equation}\label{Tdualtr}
d\tPhi=\sin^2\Theta * d\Phi  \ .
\end{equation}
While the original action was $SO(3)$
invariant the manifest symmetry of the
T-dual action (\ref{STdual}) is simply
$U(1)$ shift of $\tPhi$. As was argued
in \cite{Ricci:2007eq}  the full
$SO(3)$ symmetry group is hidden and it
is realized non-locally.

As usual in Hamiltonian formalism we
firstly  determine  from the action
(\ref{STdual}) the momenta $P_{\Theta}
\ , P_{\tPhi}$ conjugate to $\Theta,\tPhi$
\begin{eqnarray}\label{cPT}
P_\Theta&=&-\sqrt{-\gamma}\gamma^{\tau\alpha}\partial_\alpha
\Theta \ , \quad
P_{\tPhi}=-\frac{1}{\sin^2\Theta}\sqrt{-\gamma}\gamma^{\tau\alpha}
\partial_\alpha \tPhi \  \nonumber \\
\end{eqnarray}
with corresponding   Poisson brackets
\begin{equation}\label{PBS2}
\pb{\tPhi(\sigma),P_{\tPhi}(\sigma')}=
\delta(\sigma-\sigma') \ , \quad
\pb{\Theta(\sigma),P_{\Theta}(\sigma')}=
\delta (\sigma-\sigma') \ .
\end{equation}
To proceed further we have to say few
words considering the problem how Lax
connection behaves under T-duality
transformation. It is well known that
T-duality transformation
(\ref{Tdualtr}) cannot be directly
performed on the currents
 (\ref{flatorg}) and (\ref{curorg})
since they depend not only on $d\Phi$
but also explicitly on coordinate
$\Phi$. Solution of this problem is
based on the observation that for any
$g\in G$ the new current
\begin{equation}
J'=g^{-1}Jg+g^{-1}dg
\end{equation}
is again flat \footnote{We will discuss
the general procedure in section
(\ref{gener}).}. Then there exists an
element $g\in SO(3)$ that transforms
the original currents into  new ones
that depend on $\Phi$  only through its
derivatives. The forms of the matrix
$g$ and corresponding current $J'$ were
found in \cite{Ricci:2007eq} with the
result
\begin{eqnarray}
J'^1&=&-\frac{1}{\sqrt{2}} \sin\Theta
\cos\Theta (a d\Phi+b*d\Phi) \ ,
\nonumber \\
J'^2&=& -\frac{1}{\sqrt{2}} (a d\Theta+
b*d\Theta) \ , \nonumber \\
J'^3 &=&-\frac{1}{\sqrt{2}}
\sin^2\Theta (a d\Phi+b* d\Phi)+
\sqrt{2}d\Phi \ . \nonumber \\
\end{eqnarray}
Then  using (\ref{Tdualtr}) we define
T-dual flat
currents as $\hJ^A=J'^A(\Phi\rightarrow
\tPhi)$. Explicitly, their spatial
components take the form
\begin{eqnarray}
\tJ^1_{\sigma}&=&-\frac{1}{\sqrt{2}}
\frac{\cos\Theta}{\sin\Theta}
(-a\sqrt{-\gamma}\gamma^{\tau\alpha}\partial_\alpha\tPhi
+b\partial_\sigma \tPhi) \ , \nonumber \\
\tJ^2_{\sigma}&=&-\frac{1}{\sqrt{2}}
(a\partial_\sigma
\Theta-b\sqrt{-\gamma}\gamma^{\tau\alpha}
\partial_\alpha \Theta) \ , \nonumber \\
\tJ^3_{\sigma}&=&-\frac{1}{\sqrt{2}}
(b\partial_\sigma
\tPhi-a\sqrt{-\gamma}\gamma^{\tau\alpha}
\partial_\alpha \tPhi)-\frac{\sqrt{2}}{\sin^2\Theta}
\sqrt{-\gamma}\gamma^{\tau\beta}\partial_\beta
\tPhi \
\nonumber \\
\end{eqnarray}
or alternatively, as  functions of
phase space variables
$(\Theta,P_\Theta, \tPhi,P_{\tPhi})$
\begin{eqnarray}
\tJ^1_{\sigma}&=&-\frac{1}{\sqrt{2}}
a\cos\Theta\sin\Theta
P_{\tPhi}+\frac{1}{\sqrt{2}}b
\frac{\cos\Theta}{\sin\Theta}\partial_\sigma
\tPhi \ , \nonumber \\
\tJ^2_{\sigma}&=&-\frac{1}{\sqrt{2}}
a\partial_\sigma
\Theta+\frac{1}{\sqrt{2}}b P_\Theta \ ,
\nonumber \\
\tJ^3_{\sigma}&=&-\frac{1}{\sqrt{2}}
b\partial_\sigma
\tPhi+\frac{a}{\sqrt{2}} \sin^2\Theta
P_{\tPhi}+\sqrt{2}P_{\tPhi} \ . \nonumber \\
\end{eqnarray}
Then  using  (\ref{PBS2})
we  calculate the Poisson brackets of
spatial components of Lax connections
for two spectral parameters
$\Lambda,\Lambda'$.
After some straightforward calculations
we obtain
\begin{eqnarray}
 \pb{\hJ^A_\sigma(\sigma,\Lambda),
\hJ^B_\sigma(\sigma',\Lambda')}&=&
\bA^{AB}(\sigma,\Lambda,\Lambda')\delta(\sigma-\sigma')+
\bB^{AB}(\sigma,\sigma',\Lambda,\Lambda')\partial_{\sigma'}
\delta(\sigma-\sigma')+\nonumber \\
&+&\bC^{AB}(\sigma,\sigma',\Lambda,\Lambda')
\partial_\sigma \delta(\sigma-\sigma')
\ ,
\nonumber \\
\end{eqnarray}
where
\begin{eqnarray}
\bA_{\alpha\gamma,\beta\delta}(\sigma,\Lambda,\Lambda')
&=&
\bA^{AB}(\sigma,\Lambda,\Lambda')
(T_A)_{\alpha\beta}(T_B)_{\gamma\delta}
=\nonumber \\
&=&-
\frac{1}{2}[P_{\tPhi}ab'(\sin^2\Theta-\cos^2\Theta)
+bb'\frac{1}{\sin^2\Theta}\partial_\sigma
\tPhi](T_1)_{\alpha\beta}(T_2)_{\gamma\delta}+\nonumber
\\
&+&\frac{1}{2}[P_{\tPhi}
ba'(\sin^2\Theta-\cos^2\Theta)
+bb'\frac{1}{\sin^2\Theta}\partial_\sigma
\tPhi](T_2)_{\alpha\beta}(T_1)_{\gamma\delta}-\nonumber
\\
&-& ba'\sin\Theta\cos\Theta
(T_2)_{\alpha\beta}(T_3)_{\gamma\delta}+
ab'\sin\Theta\cos\Theta
(T_3)_{\alpha\beta}(T_2)_{\gamma\delta}
\ ,
\nonumber \\
\end{eqnarray}
\begin{eqnarray}
\bC_{\alpha\gamma ,\beta\delta}(\sigma,\sigma',
\Lambda,\Lambda')&=&\bC^{AB}(\sigma,\sigma',
\Lambda,\Lambda')(T_A)_{\alpha\beta}(T_B)_{\gamma\delta}=
\nonumber \\
&=&\frac{1}{2}ba'\frac{\cos\Theta(\sigma)}
{\sin\Theta(\sigma)}\cos\Theta
(\sigma')
\sin\Theta(\sigma')(T_1)_{\alpha\beta}(T_1)_{\gamma\delta}+
\nonumber \\
&+&\left(\frac{1}{2}ba'\frac{\cos\Theta(\sigma)}
{\sin\Theta(\sigma)}\sin^2\Theta(\sigma')-b\frac{\cos\Theta(\sigma)}
{\sin\Theta(\sigma)}\right)(T_1)_{\alpha\beta}(T_3)_{\gamma\delta}+
\nonumber \\
&+&\frac{1}{2}ab'
(T_2)_{\alpha\beta}(T_2)_{\gamma\delta}
+\frac{1}{2}ba'\cos\Theta(\sigma')
\sin\Theta(\sigma')
(T_3)_{\alpha\beta}(T_1)_{\gamma\delta}+
\nonumber \\
&+&\left(\frac{1}{2}ba'\sin^2\Theta(\sigma')-b\right)
(T_3)_{\alpha\beta}(T_3)_{\gamma\delta}
\nonumber \\
\end{eqnarray}
and
\begin{eqnarray}
\bB_{\alpha\gamma,\beta\delta}(\sigma,\sigma',
\Lambda,\Lambda')&=&\bB^{AB}(\sigma,\sigma',
\Lambda,\Lambda')(T_A)_{\alpha\beta}(T_B)_{\gamma\delta}=
\nonumber \\
&=& -\frac{1}{2}
ab'\cos\Theta(\sigma)\sin\Theta(\sigma)
\frac{\cos\Theta(\sigma')}{\sin\Theta(\sigma')}
(T_1)_{\alpha\beta}(T_1)_{\gamma\delta}-\nonumber
\\
&-&\frac{1}{2}ab'\cos\Theta(\sigma)\sin\Theta(\sigma)
(T_1)_{\alpha\beta}(T_3)_{\gamma\delta}
- \frac{1}{2}a'b
(T_2)_{\alpha\beta}(T_2)_{\gamma\delta}
+\nonumber
\\
&+&\left(-\frac{1}{2}ab'\sin^2\Theta(\sigma)
\frac{\cos\Theta(\sigma')}{\sin\Theta
(\sigma')}+b'\frac{\cos\Theta(
\sigma')}{\sin\Theta(\sigma')}\right)
(T_3)_{\alpha\beta}(T_1)_{\gamma\delta}+\nonumber
\\ &+&\left(-\frac{1}{2}ab'\sin^2\Theta(\sigma)
+b'\right)(T_3)_{\alpha\beta}(T_3)_{\gamma\delta}
\ .
 \nonumber \\
\end{eqnarray}
As a check, note that
$\bA^{AB}(\sigma,\Lambda,\Lambda'),
\bC^{AB}(\sigma,\sigma',\Lambda,\Lambda')$
and
$\bB^{AB}(\sigma,\sigma',\Lambda,\Lambda')$
obey the consistency relations
(\ref{dendefbAA}). Using these results
we can partially  determine the
matrices $r,s$
\begin{eqnarray}
s_{\alpha\gamma,\beta\delta}(\sigma,\Lambda,\Lambda')&=&
\frac{1}{2}(\bB_{\alpha\gamma,\beta\delta}
(\sigma,\sigma,\Lambda,\Lambda')-
\bC_{\alpha\gamma,\beta\delta}(\sigma,\sigma,\Lambda,\Lambda'))=
\nonumber \\
&=& -\frac{1}{4}
\left[(ab'+ba')\cos^2\Theta(T_1)_{\alpha\beta}
(T_1)_{\gamma\delta}
+(a'b+ba')(T_2)_{\alpha\beta}(T_2)_{\gamma\delta}
+\right.
\nonumber \\
&+&\left((ab'+ba') \cos\Theta\sin\Theta
-2b\frac{\cos\Theta}{\sin\Theta}\right)
(T_1)_{\alpha\beta}(T_3)_{\gamma\delta}+ \nonumber \\
&+&
\left((ab'+a'b)\cos\Theta\sin\Theta-2b'\frac{\cos\Theta}
{\sin\Theta}\right) (T_3)_{\alpha\beta}
(T_1)_{\gamma\delta}
+\nonumber \\
&+&\left.\left((ab'+ba')\sin^2\Theta-2(b'+b')\right)
(T_3)_{\alpha\beta}(T_3)_{\gamma\delta}\right] \nonumber \\
\end{eqnarray}
and
\begin{eqnarray}
r_{\alpha\gamma,\beta\delta}(\sigma,\Lambda,\Lambda')
&=&\frac{1}{2}[\bB_{\alpha\gamma,\beta\delta}(
\sigma,\sigma,\Lambda,\Lambda')+\bC_{\alpha\gamma,
\beta\delta}(\sigma,
\sigma,\Lambda',\Lambda)]+\hat{r}_{\alpha\gamma,\beta\delta}
(\sigma,w,v)= \nonumber \\
&=& -\frac{1}{4}
\left[(ab'-ba')\cos^2\Theta
(T_1)_{\alpha\beta}(T_1)_{\gamma\delta}
+ (ab'-ba')\cos\Theta\sin\Theta
(T_1)_{\alpha\beta}(T_3)_{\gamma\delta}-\right.
\nonumber \\
&-& (ab'-a'b)\cos\Theta\sin\Theta
(T_3)_{\alpha\beta}(T_1)_{\gamma\delta}
+(a'b-ba')
(T_2)_{\alpha\beta}(T_2)_{\gamma\delta}
+\nonumber \\
&+& \left.
\left((ab'-ba')\sin^2\Theta-2(b'-b)\right)
(T_3)_{\alpha\beta}(T_3)_{\gamma\delta}\right]+\nonumber \\
&+&
\hr^{AB}(\sigma,\Lambda,\Lambda')(T_A)_{\alpha\beta}
(T_B)_{\gamma\delta} \ ,  \nonumber \\
\end{eqnarray}
where $\hr^{AB}$ is solution of the
differential equation (\ref{hatre}).
Unfortunately, due to the fact that
$\bA,\bB,\bC$ explicitly depend on the
phase space variables it is very
difficult  to solve this differential
equation (\ref{hatre}) and we were not
able to find explicit form of
$\hr^{AB}$. On the other hand it is
important to stress that the Poisson
brackets of Lax connections take the
form as in (\ref{pbmA2}) and hence
following arguments given in appendix
we can argue that T-dual sigma model
contains an infinite number of
conserved charges that are in
involution in the sense that their
Poisson brackets vanish. In summary,
T-dual theory is classically integrable
as well in spite of the fact that the
Poisson bracket structure is intricate.
\section{Second Example: T-dual $AdS_2$ String}\label{fourth}
As the second example of T-dual theory
we consider the case of bosonic sigma
model on $AdS_2$ and its T-dual
version. Recall that the dynamics of
bosonic string on $AdS_2$ is governed
by an action
\begin{eqnarray}\label{actAd}
S
&=&-\frac{1}{2}\int d^2\sigma
\frac{1}{Y^2}\sqrt{-\gamma}
\gamma^{\alpha\beta} (\partial_\alpha X
\partial_\beta X+
\partial_\alpha Y
\partial_\beta Y) \ .
\nonumber \\
\end{eqnarray}
With analogy with previous section we
introduce three currents
\begin{eqnarray}\label{curAd}
j^1_{\alpha}&=&\frac{1}{2\sqrt{2}Y^2}
((1+(X^2-Y^2))\partial_\alpha X
+2XY\partial_\alpha Y) \ , \nonumber \\
j^2_{\alpha}&=& \frac{1}{2\sqrt{2}Y^2}
((1-(X^2-Y^2))\partial_\alpha X
-2XY \partial_\alpha Y) \ , \nonumber \\
j^3_{\alpha}&=&-\frac{1}{\sqrt{2}Y^2}
(X\partial_\alpha X+Y\partial_\alpha Y)
\nonumber \\
\end{eqnarray}
that  are conserved
\begin{equation}
\partial_\alpha [\sqrt{-\gamma}
\gamma^{\alpha\beta}j_{\beta A}]=0 \ .
A=1,2,3 \ .
\end{equation}
Further, it can be shown that these
currents are flat
\begin{equation}
\partial_\alpha j^A_\beta-
\partial_\beta j^A_\alpha+j^B_\alpha  j^C_\beta f_{BC}^{ \quad A}=0
 \ ,
\end{equation}
where $f_{BC}^{ \quad
A}=-\frac{1}{\sqrt{2}}\epsilon_{BCD}K^{DA}$
and where  the Cartan-Killing form
$K^{AB}$ is equal to $K^{AB}=
\mathrm{diag}(-1,1,1)$. Then it is easy
to see that  the sigma model action
(\ref{actAd}) can be expressed as
principal chiral model with corresponding
Lax connection
\begin{equation}\label{Laxor}
J=aj+b * j \ , \quad a=\frac{1}{2}[1\pm
\cosh \Lambda] \ , \quad b=\frac{1}{2}
\sinh \Lambda \ .
\end{equation}
Our goal is to develop Hamiltonian
formalism for T-dual theory where
T-duality is performed
along $X$ direction \cite{Ricci:2007eq}
so that  T-dual  action takes the form
\begin{equation}\label{actT2}
S=-\frac{1}{2}\int d^2\sigma
\sqrt{-\gamma} [\frac{1}{\tY^2}
\gamma^{\alpha\beta}\partial_\alpha \tX
\partial_\beta \tX+\frac{1}{\tY^2}
\gamma^{\alpha\beta}\partial_\alpha \tY
\partial_\beta \tY] \ ,
\end{equation}
where we also introduced $\tY$ defined
as
\begin{equation}
\tY=\frac{1}{Y} \ .
\end{equation}
It is clear that the action
(\ref{actT2}) again describes dynamics
of string on $AdS_2$ background and
hence the Lax connection for given
theory is the same as the original one
(\ref{Laxor}) when we replace $X,Y$
with $\tX$ and $\tY$. On the other hand
there exists Lax connection in T-dual
background that is related to the
original Lax connection by gauge
transformations and then by
substitutions $X,Y\rightarrow \tX,\tY$.
This Lax connection
was derived  in \cite{Ricci:2007eq}
and  takes the form
\begin{eqnarray}
\hJ^1_{\alpha }&=& -\frac{1}{2\sqrt{2}}
\frac{(1-\tY^2)}{\tY^2}(-a\partial_\gamma\tX
\sqrt{-\gamma}\gamma^{\gamma\delta}\epsilon_{\delta\alpha}+b\partial_\alpha
\tX)-
\sqrt{2}\frac{1}{\tY^2}\partial_\gamma
\tX
\sqrt{-\gamma}\gamma^{\gamma\delta}\epsilon_{\delta\alpha}
\ ,\nonumber \\
\hJ^2_{\alpha}&=& \frac{1}{\tY\sqrt{2}}
(a\partial_\alpha \tY-b
\partial_\gamma \tY \sqrt{-\gamma}\gamma^{\gamma\delta}
\epsilon_{\delta \alpha}) \ , \nonumber
\\
\hJ^3_{\alpha }&=& -\frac{1}{2\sqrt{2}}
\frac{(1+\tY^2)}{\tY^2}
(a \partial_\gamma \tX
\sqrt{-\gamma}\gamma^{\gamma\delta} \epsilon_{\delta
\alpha}-b
\partial_\alpha \tX)+\sqrt{2}
\frac{1}{\tY^2}\partial_\gamma
\tX \sqrt{-\gamma}\gamma^{\gamma\delta}\epsilon_{\delta\alpha}
\ .
\nonumber \\
\end{eqnarray}
To proceed further we derive from
(\ref{actT2}) the conjugate momenta
\begin{eqnarray}
P_{\tX}&=&-\frac{1}{\tY^2}\sqrt{-\gamma}\gamma^{\tau\alpha}\partial_\alpha
\tX \ , \quad
P_{\tY}=-\frac{1}{\tY^2}\sqrt{-\gamma}\gamma^{\tau\alpha}
\partial_\alpha \tY \ . \nonumber \\
\end{eqnarray}
Then the spatial components of Lax
connection expressed as  functions of
canonical variables are equal to
\begin{eqnarray}
\hJ^{1}_\sigma&=&-\frac{1}{2\sqrt{2}}
\frac{(1-\tY^2)}{\tY^2}(a\tY^2 P_{\tX}
+b\partial_\sigma \tX)+
\sqrt{2}P_{\tX}\ , \nonumber \\
\hJ^2_\sigma &=& \frac{1}{\sqrt{2}\tY}
(a\partial_\sigma \tY+b \tY^2P_{\tY}) \
,  \nonumber
\\
\hJ^{3}_\sigma &=& -\frac{1}{2\sqrt{2}}
\frac{(1+\tY^2)}{\tY^2} (a \tY^2
P_{\tX} +b
\partial_\sigma \tX)-\sqrt{2}P_{\tX} \ . \nonumber\\
\end{eqnarray}
Now we are ready to determine Poisson
bracket of spatial components
of Lax connection. Again, after some
calculations we derive the Poisson
brackets that have the same form as
in (\ref{pbmA1})
where the matrices $\bA,\bB$ and
$\bC$ are equal to
\begin{eqnarray}\label{bAA}
\bA_{\alpha\gamma,\beta\delta}(\sigma,
\Lambda,\Lambda')&=&[\frac{b'}{2\tY^2}
(a\tY^2 P_{\tX}+b\partial_\sigma\tX)
-\frac{1}{2}ab'(1-\tY^2)\tY^2 P_{\tX}
](T^1)_{\alpha\beta}(T^2)_{\gamma\delta}+ \nonumber \\
\nonumber \\
&+&[-\frac{b}{2\tY^2}(a'\tY^2
P_{\tX}+b'
\partial_\sigma \tX)
+\frac{1}{2}ba'(1-\tY^2)\tY^2
P_{\tX}](T^2)_{\alpha\beta}
(T^1)_{\gamma\delta}+
 \nonumber \\
&+&\frac{1}{2\tY}
(ba'-ab')\partial_\sigma \tY
(T^2)_{\alpha\beta}(T^2)_{\gamma\delta}+\nonumber \\
&+&[\frac{b}{2\tY^2}(a'\tY^2 P_{\tX}+b'\partial_\sigma
\tX)
+\frac{1}{2}ba'(1+\tY^2)P_{\tX}](T^2)_{\alpha\beta}
(T^3)_{\gamma\delta}+
\nonumber \\
&-&[\frac{b'}{2\tY^2} (a\tY^2 P_{\tX}+b
\partial_\sigma \tX)
+\frac{1}{2}ab'(1+\tY^2) P_{\tX}
](T^3)_{\alpha\beta}(T^2)_{\gamma\delta}
\nonumber \\
 \end{eqnarray}
and
\begin{eqnarray}\label{bCA}
\bC_{\alpha\gamma,\beta\delta}(\sigma,
\sigma',\Lambda,\Lambda')&=& \nonumber \\
&=&[ \frac{ba'}{8}
\frac{(1-\tY^2(\sigma))}{\tY^2(\sigma)}
(1-Y^2(\sigma'))- \frac{b}{2}
\frac{(1-\tY^2(\sigma))}{\tY^2(\sigma)}]
(T^1)_{\alpha\beta}(T^1)_{\gamma\delta}+
\nonumber \\
&+&
[\frac{ba'}{8}\frac{(1-\tY^2(\sigma))}
{\tY^2(\sigma)}(1+\tY^2(\sigma'))
+\frac{b}{2}\frac{(1-\tY^2(\sigma))}{\tY^2(\sigma)}
](T^1)_{\alpha\beta}(T^3)_{\gamma\delta}
+\nonumber \\
&+&\frac{ab'}{2}\frac{\tY^2(\sigma')}
{\tY(\sigma)}(T^2)_{\alpha\beta}
(T^2)_{\gamma\delta}+\nonumber \\
&+&[\frac{ba'}{8}
\frac{(1+\tY^2(\sigma))}{\tY^2(\sigma)}
(1-\tY^2(\sigma'))
-\frac{b}{2}\frac{1+\tY^2(\sigma)}{\tY^2(\sigma)}]
(T^3)_{\alpha\beta}(T^1)_{\gamma\delta}+\nonumber
\\
&+&[\frac{ba'}{8}\frac{(1+\tY^2(\sigma))}
{\tY^2(\sigma)}(1+\tY^2(\sigma'))
-\frac{b}{2}\frac{(1+\tY^2(\sigma))}{\tY^2(\sigma)}]
(T^3)_{\alpha\beta}(T^3)_{\gamma\delta}
\ ,
\nonumber \\
\bB_{\alpha\gamma,\beta\delta}
(\sigma,\sigma',\Lambda,
\Lambda')&=&-[\frac{ab'}{8}
(1-\tY^2(\sigma)\frac{(1-\tY^2(\sigma'))}{
\tY^2(\sigma')}
-\frac{1}{2}\frac{(1-\tY^2(\sigma'))}
{\tY^2(\sigma')}b'](T^1)_{\alpha\beta}
(T^1)_{\gamma\delta}- \nonumber \\
&-&[\frac{ab'}{8}
(1-\tY^2(\sigma))\frac{(1+\tY^2(\sigma'))}{\tY^2(\sigma')}
-\frac{b'}{2}\frac{(1+\tY^2(\sigma'))}
{\tY^2(\sigma')}](T^1)_{\alpha\beta}(T^3)_{\gamma\delta}
-\nonumber \\
&-&\frac{ba'}{2}\frac{\tY^2(\sigma)}{\tY(\sigma')}
(T^2)_{\alpha\beta}(T^2)_{\gamma\delta}-\nonumber
\\
&-&[\frac{ab'}{8}(1+\tY^2(\sigma))
\frac{(1-\tY^2(\sigma'))}{\tY^2(\sigma')}
+\frac{b'}{2}\frac{(1-\tY^2(\sigma'))}{\tY^2(\sigma')}]
(T^3)_{\alpha\beta}(T^1)_{\gamma\delta}
-\nonumber \\
&-&[\frac{a'b}{8}(1+\tY^2(\sigma))
\frac{(1+\tY^2(\sigma'))}
{\tY^2(\sigma')}
-\frac{b'}{2}\frac{(1+\tY^2(\sigma'))}{\tY^2(\sigma')}]
(T^3)_{\alpha\beta}(T^3)_{\gamma\delta}
\ .
\nonumber \\
\end{eqnarray}
As a check note that the matrices
(\ref{bAA}),(\ref{bCA}) obey the
consistency relations (\ref{gendefbA}).
Further, we can also determine the
matrix $s^{AB}$ however we are not able
to fully determine $r^{AB}$ due to the
fact that the matrices $\bA,\bB,\bC$
are functions of phase space variables.
It is clear that the theory is
classically integrable since we can in
principle find an infinite number of
charges that are in involutions.
However the consequence of the
non-local nature of T-dual Lax
connection is that the matrices
$r,s$ now explicitly depend on phase
space variables and are non-local. On
the other hand the case of $AdS_2$
 is exceptional since we know that
its T-dual image is again $AdS_2$ so
that we can find Lax connection
corresponding standard principal chiral
with constant $r$ and $s$ matrices.
 We are not going to study the
relations between these two Lax
connections in this paper. We hope
to return to the study of this
problem in  future publication.


\section{General
procedure}\label{gener}
 In this section we consider  general
 situation when we have principal
 chiral model with a field $g(\sigma)$
 that maps the string world-sheet into
 some group $G$ with Lie algebra
 $\mathbf{g}$. Further we presume
that the Lie
 algebra $\mathbf{g}$   has generators $T_A,
 A=1,\dots,\mathrm{dim}(\mathbf{g})$
 that obey the  relation
 \begin{equation}
 [T_A,T_B]=f_{AB}^{\quad C}T_C \ .
 \end{equation}
 From $g(x)$ we can construct
 a current $j$ in the form
 \begin{equation}
 j=g^{-1}dg\equiv
 E^A_M dx^MT_A \ ,
 \end{equation}
where by definition
\begin{equation}
dj+j\wedge j=0 \ ,
\end{equation}
and where we introduced sigma model
coordinates $x^M$. Then the dynamics of
the theory is governed by the action
\begin{equation}
S=-\frac{1}{2}\int d^2\sigma
\sqrt{-\gamma}\gamma^{\alpha\beta}
K_{AB}E^A_M \partial_\alpha x^M E^B_N
\partial_\beta x^N \ ,
\end{equation}
where $K_{AB}=\tr (T_A T_B)$.  As we
reviewed in section (\ref{second}) the
principal chiral model possesses Lax
connection $J=aj+b*j$
 that is flat
\begin{equation}
 dJ+J\wedge J=0 \
 \end{equation}
 for $a=\frac{1}{2}[1\pm \cosh\Lambda] \ ,
b=\frac{1}{2} \sinh\Lambda$, where $\Lambda$
is a spectral parameter.

 The principal chiral model
 has an important property
that when we perform
 the gauge
transformation from $g\in G$ on the
original \emph{Lax connection}
\begin{equation}\label{Jc'}
J'=g^{-1}Jg+g^{-1}dg
\end{equation}
we obtain that the new one is again
flat
\begin{eqnarray}
dJ'+J'\wedge J'
=g^{-1}(dJ+J\wedge J)g=0 \ . \nonumber \\
\end{eqnarray}
To proceed further  we write the gauge
transformation (\ref{Jc'}) in component
formalism. Since $J'=J'^AT_A$ we obtain
\begin{eqnarray}\label{J'com}
J'^A=J^C \Omega_C^{\ A}+e^A \ , \nonumber \\
\end{eqnarray}
where
 \begin{equation}\label{defg}
g^{-1}dg=e^AT_A \ , \quad
 \Omega_{CB}=\tr
(g^{-1}T_C gT_B)\ , \quad \Omega_C^{\
A}=\Omega_{CB}K^{BA} \ ,
\end{equation}
where generally  $\Omega_C^{\ A} \ ,
e^A$ are functions of phase space
variables.
Our goal is to determine the Poisson
bracket of Lax connection in T-dual
theory. The
first step in this direction is to
determine the Poisson bracket of Lax
connection $J'$. Using (\ref{J'com}) we
obtain
\begin{eqnarray}\label{J'AB}
\left\{J'^A_\sigma(\sigma,\Lambda)\right.
&,& \left. J'^B_\sigma
(\sigma',\Gamma)\right\}=\pb{e^A_\sigma(\sigma),e^B_\sigma(\sigma')}+
\pb{e^A_\sigma(\sigma),J^C_\sigma
(\Gamma,\sigma')}\Omega_C^{\ B}(\sigma')+
\nonumber \\
&+&\pb{e^A_\sigma(\sigma),\Omega_C^{ \ B}(\sigma')}J^C_\sigma(\Gamma,\sigma')+
\pb{J^C_\sigma(\Lambda,\sigma),e^B_\sigma(\sigma')}\Omega_C^{ \ A}(\sigma)+
\nonumber \\
&+&J^C_\sigma(\Lambda,\sigma)\pb{\Omega_C^{ \ A}(\sigma),e^B_\sigma(\sigma')}+
\Omega_C^{ \ A}(\sigma)\pb{J^C_\sigma(\Lambda,\sigma),J^D_\sigma(\Gamma,\sigma')}
\Omega_D^{ \ B}(\sigma')+\nonumber \\
&+&\pb{J^C_\sigma(\Lambda,\sigma),\Omega_D^{ \ B}(\sigma')}
\Omega_C^{ \ A}(\sigma)J^D_\sigma
(\Gamma,\sigma')+
\nonumber \\
&+&J^C_\sigma(\Lambda,\sigma)\pb{\Omega_C^{\ A}(\sigma),J^D_\sigma(\Gamma,\sigma')}
\Omega_D^{ \ B}(\sigma')+
J^C_\sigma(\sigma,\Lambda)\pb{\Omega_C^{ \ A}(\sigma),
\Omega_D^{ \ B}(\sigma')}J^B_\sigma(\sigma',\Gamma) \ .
\nonumber \\
\end{eqnarray}
Let us now presume that $g$ is function
of $x^M$ only. Then the spatial component
$e^A_{\sigma}$
depends on $x^M$ and their
 derivatives  $x^M$
and does not depend on $p_M$. It
is also clear that
$\Omega_A^{\ B}$ depends on $x^M$ only. Then
we obtain
\begin{eqnarray}
\pb{e^A_\sigma(\sigma),e^B_\sigma(\sigma')}=0  \ ,
\quad \pb{e^A_\sigma(\sigma),\Omega_B^{ \
C}(\sigma')}=0 \ ,
\quad \pb{\Omega_A^{ \ B}(\sigma),
\Omega_C^{ \ D}(\sigma')}=0 \ .
\nonumber \\
\end{eqnarray}
Then, using the arguments
above and  due to the fact that $J^C_\sigma$ is linear
in momenta  we can presume that
\begin{eqnarray}
\pb{e^A_\sigma(\sigma),J^B_\sigma(\sigma',\Lambda)}&=&
\mE^{AB}(\sigma,\sigma',\Lambda)\partial_\sigma
\delta(\sigma-\sigma')+\nonumber \\
&+&
\mF^{AB}(\sigma,\sigma',\Lambda)
\partial_{\sigma'}\delta(\sigma-\sigma')+
\mG^{AB}
(\sigma,\Lambda)\delta(\sigma-\sigma') \ ,
\nonumber \\
\pb{\Omega_A^{\ B}(\sigma),
J^C_\sigma(\Gamma,\sigma')}&=& \tOmega_{A}^{\ BC}
(\sigma,\Gamma)\delta(\sigma-\sigma')
\ . \nonumber \\
\end{eqnarray}
Further, let us presume that
Poisson brackets between
$J^A_\sigma(\sigma,\Lambda)$ and
$J^B_\sigma(\sigma',\Gamma)$ take the
form
\begin{eqnarray}\label{pbJABC}
\left\{J^A_\sigma
(\sigma,\Lambda)\right. &,&\left.
J^B_\sigma(\sigma',\Gamma)\right\}=
\bA^{AB}(\sigma,\Lambda,\Gamma)\delta(\sigma-\sigma')+\nonumber \\
&+&\bB^{AB}(\sigma,\sigma',\Lambda,\Gamma)\partial_{\sigma'}
\delta(\sigma-\sigma')+
\bC^{AB}(\sigma,\sigma',\Lambda,\Gamma)
\partial_\sigma \delta(\sigma-\sigma')
\ .
\nonumber \\
\end{eqnarray}
Then  if we insert this  expression into
(\ref{J'AB}) we obtain
that the Poisson bracket of Lax connections
$J'$ has the same form as (\ref{pbJABC})
\begin{eqnarray}\label{J'ABC}
\pb{J'^A_\sigma(\sigma,\Lambda),J'^B_\sigma
(\sigma',\Gamma)}
&=& \bA'^{AB}(\sigma,\Lambda,\Gamma)
\delta(\sigma-\sigma')+\nonumber \\
&+& \bC'^{AB}
(\sigma,\sigma',\Lambda,\Gamma)
\partial_{\sigma}\delta(\sigma-\sigma')+
\bB'^{AB}(\sigma,\sigma',\Lambda,\Gamma)
\partial_{\sigma'}\delta(\sigma-\sigma') \ ,
\nonumber \\
\end{eqnarray}
where
\begin{eqnarray}\label{ABCgeneral}
\bA'^{AB}(\sigma,\Lambda,\Gamma)&=&
\mG^{AC}(\sigma,\Gamma)\Omega_C^{ \
B}(\sigma)- \Omega_C^{ \
A}(\sigma)\mG^{BC}(\sigma,\Lambda)+\nonumber \\
&+&
\Omega_C^{ \
A}(\sigma)\bA^{CD}(\sigma,\Lambda,\Gamma)
\Omega_D^{ \ B}(\sigma)-
J^D_\sigma
(\Gamma,\sigma)
\tOmega_{D}^{\ BC}(\sigma,\Lambda)\Omega_C^{
\ A}(\sigma)+\nonumber \\
&+&
J^C_\sigma(\Lambda,\sigma)\tOmega^{\ DA}_{C}
 (\sigma,\Gamma)\Omega_D^{ \
B}(\sigma)
 \ , \nonumber \\
 \bB'^{AB}(\sigma,\sigma',\Lambda,\Gamma)&=&
 \mF^{AC}(\sigma,\sigma',\Gamma)\Omega_C^{ \ B}(\sigma')-\nonumber \\
 &-&
 \Omega_C^{ \ A}(\sigma)\mE^{BC}(\sigma',\sigma,\Lambda)+
  \Omega_C^{ \ A}(\sigma)\bB^{CD}(\sigma,\sigma',\Lambda,\Gamma)
 \Omega_D^{ \ B}(\sigma') \ ,
 \nonumber \\
 \bC'^{AB}(\sigma,\sigma',\Lambda,\Gamma)&=&
 \mE^{AC}(\sigma,\sigma',\Gamma)\Omega_C^{ \ B}(\sigma')
 -\nonumber \\
&-&\Omega_C^{ \ A}(\sigma)\mF^{BC}(\sigma',\sigma,\Lambda)
 +\Omega_C^{ \ A}(\sigma)\bC^{CD}(\sigma,\sigma',\Lambda,\Gamma)
 \Omega_D^{ \ B}(\sigma') \ .  \nonumber \\
\end{eqnarray}
Now we are ready to calculate the
Poisson brackets of Lax connection in
T-dual theory.  Let us denote
$\eta^I(\sigma)=
\left(x^1(\sigma),\dots,x^n(\sigma),
p_1(\sigma),\dots,p_n(\sigma)\right) \
, I=1,\dots,2n$ where $n$ is dimension
of $M$  and introduce
symplectic structure
$\bJ^{IJ}(\sigma,\sigma')$ defined as
\begin{equation}
\bJ=\left(\begin{array}{cc} 0 &
I_{n\times n}\delta(\sigma-\sigma') \\
-I_{n\times n}\delta(\sigma-\sigma') & 0 \\
\end{array}\right) \ .
\end{equation}
Then   the Poisson bracket of two
functions $F(\eta),G(\eta)$ can be
written as
\begin{eqnarray}
\pb{F,G}_\eta= \int d\sigma d\sigma'
\left(\frac{\delta F}{\delta
\eta^I(\sigma)}\bJ^{IJ}(\sigma,\sigma')\frac{\delta
G}{\delta \eta^J(\sigma')}\right) \ .
\nonumber \\
\end{eqnarray}
For example, if
$F=x^N(\sigma),G=p_M(\sigma')$ we
obtain
\begin{equation}
\pb{x^N(\sigma),p_M(\sigma')}=
\delta^N_M\delta(\sigma-\sigma')
\end{equation}
or more covariantly
\begin{equation}
\pb{\eta^I(\sigma),
\eta^J(\sigma')}_\eta=
\bJ^{IJ}(\sigma,\sigma') \ .
\end{equation}
An important property of T-duality
 is that it is a sort of
canonical transformation. More
precisely, let us denote variables in
T-dual theory as $\teta^I=
\left(\tilde{x}^M(\sigma),\tilde{p}_M(\sigma)\right)$
and presume that they can be expressed
as functions of original variables
\begin{equation}
\teta^I(\sigma)= \teta^I(\eta(\sigma))
\ .
\end{equation}
Now if the transformation from $\eta$
to $\teta$ is canonical, we have that
the matrix
\begin{equation}
M^I_J(\sigma,\sigma')= \frac{\delta
\teta^I(\sigma)}{\delta
\eta^J(\sigma')}
\end{equation}
preserves the symplectic structure in
the sense that
\begin{equation}
\int d x dy M^I_K(\sigma, x)
\bJ^{KL}(x,y)M_L^J(y,\sigma')=
\bJ^{IJ}(\sigma,\sigma') \ .
\end{equation}
Using this fact we immediately obtain
\begin{eqnarray}\label{Tcano}
\pb{\teta^I(\sigma),\teta^J(\sigma)}_{\teta}=
\bJ^{IJ}(\sigma,\sigma) \ . \nonumber \\
\end{eqnarray}
This expression implies that   all Poisson brackets are
invariant under canonical
transformations. In fact, let us
consider two functionals
$F(\eta)$ and $G(\eta)$. Then
the invariance of Poisson bracket under
cannonical transformation (\ref{Tcano})
implies
\begin{eqnarray}
\pb{F(\eta),G(\eta)}_\eta=
\pb{F(\teta),G(\teta)}_{\teta} \ ,
\nonumber \\
\end{eqnarray}
where $\eta$ and $\teta$ are related
by cannonical transformation.
Then  if we apply these considerations to
the case of Lax connections in original
and T-dual theories we obtain
\begin{eqnarray}
\left\{\hJ^A_\sigma(\teta(\sigma),\Lambda)\right.
&,&\left.
\hJ^B_\sigma(\teta(\sigma'),\Gamma)\right\}_{\teta}=
\pb{\hJ^A_\sigma(\Lambda,\eta(\sigma)),
\hJ^B_\sigma(\eta(\sigma'),\Gamma)}_\eta=\nonumber \\
&=&
\bA'^{AB}(\eta(\sigma),\Lambda,\Gamma)
\delta(\sigma-\sigma')+
\bC'^{AB}
(\eta(\sigma),\eta(\sigma'),\Lambda,\Gamma)
\partial_{\sigma}\delta(\sigma-\sigma')+\nonumber \\
&+&\bB'^{AB}(\eta(\sigma),\eta(\sigma'),\Lambda,\Gamma)
\partial_{\sigma'}\delta(\sigma-\sigma')=
\nonumber \\
&=&\tbA^{AB}(\teta(\sigma),\Lambda,\Gamma)
\delta(\sigma-\sigma')+ \tbC^{AB}
(\teta(\sigma),\teta(\sigma'),\Lambda,\Gamma)
\partial_{\sigma}\delta(\sigma-\sigma')+
\nonumber \\
&+&\tbB^{AB}(\teta(\sigma),\teta(\sigma'),\Lambda,\Gamma)
\partial_{\sigma'}\delta(\sigma-\sigma')
\ ,
\nonumber \\
\end{eqnarray}
where
\begin{eqnarray}
\tbA^{AB}(\teta(\sigma),\Lambda,\Gamma)
&\equiv& \bA'^{AB}(\eta(\teta(\sigma)),
\Lambda,\Gamma) \ , \nonumber \\
\tbB^{AB}(\teta(\sigma),\teta(\sigma'),
\Lambda,\Gamma) &  \equiv &
\bB'^{AB}(\eta(\teta(\sigma)),\eta(\teta(\sigma')),
\Lambda,\Gamma) \ , \nonumber \\
\tbC^{AB}(\teta(\sigma),\teta(\sigma'),\Lambda,\Gamma)
&\equiv&
\bC'^{AB}(\eta(\teta(\sigma)),\eta(\teta(\sigma')),
\Lambda,\Gamma) \ , \nonumber \\
\end{eqnarray}
and where in the first step we used
(\ref{J'ABC}) and in the second
one we expressed (\ref{ABCgeneral})
as functions of $\teta$. This
result implies that we can  express the Poisson bracket
of Lax connections in T-dual theory using
the known form of  original Lax connection $J_\sigma^A$
and the known form of
 $e^A_\sigma \ , \Omega_A^{\ B}$ that
are in the final step
 expressed as functions of T-dual variables. Then is
it clear from (\ref{ABCgeneral}) that
 the dual  theory is again  classically integrable
in the sense that there are infinite
number of integrals of motion that are
in involution. On the other hand the
complicated  form of matrices
$\tilde{\bA},\tilde{\bB}, \tilde{\bC}$
implies that generally the matrices
$\tilde{s}$ and $\tilde{r}$ are
functions
 of phase space variables. Moreover
we can also expect  that  $\tilde{r}$ is
non-local due to the fact that $\tilde{\hr}$
is a solution of the differential equation
(\ref{hatre}) with tilded functions
$\tilde{\bA},\tilde{\bB},\tilde{\bC}$.

\vskip .2in \noindent {\bf
Acknowledgements:} This work was
 supported by the Czech
Ministry of Education under Contract
No. MSM 0021622409.

\begin{appendix}
\section{Appendix: Review of Basic Properties
of Monodromy Matrix}\label{appendix} In this appendix
we  review  properties of monodromy
matrix, following
\cite{Maillet:1985ek,Maillet:1985ec}.
The monodromy matrix
$\mT_{\alpha\beta}(\sigma,\sigma',\Lambda)$,
where $\Lambda$ is spectral parameter
and where $\alpha,\beta$ correspond to
matrix indices  can be defined as
\begin{eqnarray}\label{diffdefT}
\partial_\sigma \mT_{\alpha\beta}
(\sigma,\sigma',\Lambda)&=&
\mA_{\alpha\gamma}(\sigma,\Lambda)
\mT_{\gamma\beta}(\sigma,\sigma',\Lambda)
\ ,
\nonumber \\
\partial_{\sigma'}
\mT_{\alpha\beta}(\sigma,\sigma',\Lambda)&=&
-\mT_{\alpha\gamma}
(\sigma,\sigma',\Lambda)\mA_{\gamma\beta}(\sigma',\Lambda)
\
\nonumber \\
\end{eqnarray}
with the normalization condition
\begin{equation}
\mT_{\alpha\beta} (\sigma
,\sigma,\Lambda)=\delta_{\alpha\beta} \
\end{equation}
and
\begin{equation}
\mT^{-1}_{\alpha\beta}(\sigma,\sigma',\Lambda)=
\mT_{\alpha\beta}(\sigma',\sigma,\Lambda)
\ .
\end{equation}
Note that in our notation
$\mA_{\alpha\beta} (\sigma,\Lambda)$ is
the spatial component of Lax
connection.

The main interest in the theory of
integrable systems is the Poisson
bracket between $\mT(\Lambda)$ and
$\mT(\Gamma)$. As was shown in a nice
way in \cite{Maillet:1985ek} the
Poisson bracket between $
\mT_{\alpha\beta}(\sigma,\sigma',\Lambda)$
and
$\mT_{\gamma\delta}(\xi,\xi',\Gamma)$
where  all $\sigma,\sigma',\xi,\xi'$
 are distinct is equal to
\begin{eqnarray}
& &
\pb{\mT_{\alpha\beta}(\sigma,\sigma',\Lambda),
\mT_{\gamma\delta}(\xi,\xi',\Gamma)}=
\int_\sigma^{\sigma'} d\sigma_1
\int_{\xi}^{\xi'} d\xi_1
\mT_{\alpha\sigma_1}(\sigma,\sigma_1,\Lambda)
\mT_{\gamma\rho_1}(\xi,\xi_1,\Gamma)\times
\nonumber \\
&\times &
\pb{\mA_{\sigma_1\sigma_2}(\sigma_1,\Lambda),
\mA_{\rho_1\rho_2}(\xi_1,\Gamma)}
\mT_{\sigma_2\beta}(\sigma_1,\sigma',\Lambda)
\mT_{\rho_2\delta}(\xi_1,\xi',\Gamma) \
.
\nonumber \\
\end{eqnarray}
The result above suggests that
fundamental role in the theory of
integrable systems plays the Poisson
bracket between spatial components of
Lax connection. Let us now presume that
the Poisson bracket of spatial
components of Lax connection
$\mA(\sigma,\Lambda)$ and
$\mA(\sigma',\Gamma)$ takes the form
\begin{eqnarray}\label{pbmA1}
\pb{\mA_{\alpha\beta}(\sigma,\Lambda),
\mA_{\gamma\delta}(\sigma',\Gamma)}&=&
\bA_{\alpha\gamma,\beta\delta}(\sigma,\Lambda,\Gamma)
\delta(\sigma-\sigma')+\nonumber \\
&+&\bB_{\alpha\gamma,\beta\delta}(\sigma,\sigma',\Lambda,\Gamma)
\partial_{\sigma'} \delta(\sigma-\sigma')+
\bC_{\alpha\gamma,\beta\delta}(\sigma,\sigma',\Lambda,\Gamma)
\partial_{\sigma}\delta(\sigma-\sigma') \
,
\nonumber \\
\end{eqnarray}
where due to the antisymmetric property
of Poisson brackets the functions
$\bA,\bC,\bB$ obey consistency
relations
\begin{eqnarray}\label{gendefbA}
\bA_{\alpha\gamma,\beta\delta}(\sigma,\Lambda,\Gamma)&=&
-\bA_{\gamma\alpha,\delta\beta}(\sigma,\Gamma,\Lambda)
\ ,
\nonumber \\
\bB_{\alpha\gamma,\beta\delta}(\sigma,\sigma',\Lambda,\Gamma)&=&
-\bC_{\gamma\alpha,\beta\delta}(\sigma',\sigma,\Gamma,\Lambda) \ ,
\nonumber \\
\bC_{\alpha\gamma,\beta\delta}(\sigma,\sigma',\Lambda,\Gamma)&=&
-\bB_{\gamma\alpha,\delta\beta}(\sigma',\sigma,\Gamma,\Lambda)
\ .
\nonumber \\
\end{eqnarray}
Further, let us presume that  the Lax
connection takes value in the Lie
algebra $\mathbf{g}$ os some group $G$.
 Let us then presume that
generators of the algebra $\mathbf{g}$
are $T_A,A=1,\dots, \mathrm{dim}(G)$
with following structure
$[T_A,T_B]=f_{AB}^{\quad C}T_C$.
 Then we can write
$\mA$ in the form
\begin{equation}
\mA
(\sigma,\Lambda)=J^A_\sigma(\sigma,\Lambda)
T_A
\end{equation}
and also
\begin{eqnarray}
\bA_{\alpha\gamma,\beta\delta}
(\sigma,\Lambda,\Gamma)&=&
\bA^{AB}(\sigma,\Lambda,\Gamma)
(T_A)_{\alpha\beta}(T_B)_{\gamma\delta}
\ ,
\nonumber \\
\bB_{\alpha\gamma,\beta\delta}
(\sigma,\sigma'\Lambda,\Gamma)&=&
\bB^{AB}(\sigma,\sigma',\Lambda,\Gamma)
(T_A)_{\alpha\beta}(T_B)_{\gamma\delta}
\ ,
\nonumber \\
\bC_{\alpha\gamma,\beta\delta}
(\sigma,\sigma',\Lambda,\Gamma)&=&
\bC^{AB}(\sigma,\sigma',\Lambda,\Gamma)
(T_A)_{\alpha\beta}(T_B)_{\gamma\delta}
\ .
\nonumber \\
\end{eqnarray}
Then the Poisson bracket
(\ref{pbmA1}) can be written as
\begin{eqnarray}\label{pbmA2}
\pb{J^A_{\sigma}(\sigma,\Lambda),
J^B_{\sigma}(\sigma',\Gamma)}&=&
\bA^{AB}(\sigma,\Lambda,\Gamma)
\delta(\sigma-\sigma')+\nonumber \\
&+&\bB^{AB}(\sigma,\sigma',\Lambda,\Gamma)
\partial_{\sigma'} \delta(\sigma-\sigma')+
\bC^{AB}(\sigma,\sigma',\Lambda,\Gamma)
\partial_{\sigma}\delta(\sigma-\sigma') \
\nonumber \\
\end{eqnarray}
and the relations (\ref{gendefbA})
take alternative form
\begin{eqnarray}\label{dendefbAA}
\bA^{AB}(\sigma,\Lambda,\Gamma)&=&
-\bA^{BA}(\sigma,\Gamma,\Lambda) \ ,
\nonumber \\
\bB^{AB}(\sigma,\sigma',\Lambda,\Gamma)&=&
-\bC^{BA}(\sigma',\sigma,\Gamma,\Lambda)
\ . \nonumber \\
\end{eqnarray}
 Let us introduce the matrices
$r_{\alpha\gamma,\beta\delta}(\sigma,\Lambda,\Gamma),
s_{\alpha\gamma,\beta\delta}(\sigma,\Lambda,\Gamma)$
defined as
\begin{eqnarray}
s_{\alpha\gamma,\beta\delta}(\sigma,\Lambda,\Gamma)
&=&\frac{1}{2}(\bB_{\alpha\gamma,\beta,\delta}
(\sigma,\sigma,\Lambda,\Gamma)-
\bC_{\alpha\gamma,\beta\delta}(\sigma,\sigma,\Lambda,\Gamma))
\nonumber \\
r_{\alpha\gamma,\beta\delta}(\sigma,\Lambda,\Gamma)
&=&\frac{1}{2}(\bB_{\alpha\gamma,\beta\delta}(
\sigma,\sigma,\Lambda,\Gamma)+\bC_{\alpha\gamma,
\beta\delta}(\sigma,
\sigma,\Lambda,\Gamma))+\hat{r}_{\alpha\gamma,\beta\delta}
(\sigma,w,v) \ ,
 \nonumber \\
\end{eqnarray}
or alternatively
\begin{eqnarray}
s^{AB}(\sigma,\Lambda,\Gamma)&=&\frac{1}{2}(
\bB^{AB}(\sigma,\sigma,\Lambda,\Gamma)-
\bC^{AB}(\sigma,\sigma,\Lambda,\Gamma))=
\nonumber \\
&=& s^{BA}(\sigma,\Gamma,\Lambda)
\nonumber \\
r^{AB}(\sigma,\Lambda,\Gamma)
&=&\frac{1}{2}(
\bB^{AB}(\sigma,\sigma,\Lambda,\Gamma)+
\bC^{AB}(\sigma,\sigma,\Lambda,\Gamma))+
\hr^{AB}(\sigma,\Lambda,\Gamma))=\nonumber
\\
&=&-r^{BA}(\sigma,\Gamma,\Lambda) \ ,
  \nonumber \\
\end{eqnarray}
where
$\hat{r}^{AB}(\sigma,\Lambda,\Gamma)$
is solution of the inhomogeneous first
order differential equation
\begin{eqnarray}\label{hatre}
\partial_\sigma \hr^{AB}+
\hr^{DB}f_{DC}^{\quad A} J^D_\sigma(\Lambda)+
\hr^{AD}f_{DC}^{\quad B}J^C_\sigma(\Gamma)=
\Omega^{AB} \ ,  \nonumber \\
\end{eqnarray}
where \begin{eqnarray}
 \Omega^{AB}
(\sigma,\Lambda,\Gamma)=
\bA^{AB}(\sigma,\Lambda,\Gamma)-
\partial_u (\bB
(\sigma,u,\Lambda,\Gamma)
+\bC(u,\sigma,\Lambda,\Gamma))
^{AB}_{u=\sigma}-
\nonumber \\
-
\bB^{AC}(\sigma,\sigma,\Lambda,\Gamma)
f_{CD}^{\quad B}J^D_\sigma(\sigma,\Gamma)
+J^C_\sigma(\sigma,\Lambda) f_{CD}^{\quad A}
\bC^{DB}(\sigma,\sigma,\Lambda,\Gamma)
\nonumber \\
\end{eqnarray}
The significance of matrices $s$
and $r$ is that with the help of
them  we can write the Poisson bracket
(\ref{pbmA1}) in the form
\cite{Maillet:1985ek}
\begin{eqnarray}\label{pbmA1a}
\left\{\mA_{\alpha\beta}(\sigma,\Lambda)\right.
& ,&
\left.\mA_{\gamma\delta}(\sigma',\Gamma)\right
\}= (\partial_\sigma
r_{\alpha\gamma,\beta\delta}
(\sigma,\Lambda,\Gamma)-
\partial_\sigma s_{\alpha\gamma,\beta\delta}
(\sigma,\Lambda,\Gamma))\delta(\sigma-\sigma')
+\nonumber \\
&+&[(r_{\alpha\gamma,\sigma\delta}(\sigma,\Lambda,\Gamma)-
s_{\alpha\gamma,\sigma \delta}
(\sigma,\Lambda,\Gamma))
\mA_{\sigma\beta}(\sigma,\Lambda)-\nonumber \\
&-&\mA_{\alpha\sigma}(\sigma,\Lambda)
(r_{\sigma\gamma,\beta\delta}(\sigma,\Lambda,\Gamma)
-s_{\sigma\gamma,\beta\delta}
(\sigma,\Lambda,\Gamma))]\delta(\sigma-\sigma')+
\nonumber \\
&+&
[(r_{\alpha\gamma,\beta\sigma}(\sigma,\Lambda,\Gamma)+
s_{\alpha\gamma,\beta\sigma}(\sigma,
\Lambda,\Gamma))
\mA_{\sigma\delta}(\Gamma,\sigma)-\nonumber \\
&-&\mA_{\gamma\sigma}(\Gamma,\sigma)(r_{\alpha\sigma,\beta\delta}
(\sigma,\Lambda,\Gamma)+s_{\alpha\sigma,\beta\delta}
(\sigma,\Lambda,\Gamma))]\delta(\sigma-\sigma')-\nonumber
\\
&-&2
s_{\alpha\gamma,\beta\delta}(\sigma,\Lambda,\Gamma)
\partial_{\sigma}\delta(\sigma-\sigma')
\nonumber \\
\end{eqnarray}
or alternatively
\begin{eqnarray}\label{pbLaxa}
\pb{J^A_\sigma(\sigma,\Lambda),J^B_\sigma(\sigma',
\Gamma)} &=&
(r-s)^{CB}(\sigma,\Lambda,\Gamma)
f_{CD}^{ \quad A} J^D_\sigma(\sigma,\Lambda)
\delta(\sigma-\sigma') +\nonumber \\
 &+&(r+s)^{AC}(\sigma,\Lambda,\Gamma)
 f_{CD}^{\quad B} J^D_\sigma(\sigma,\Gamma)
 \delta(\sigma-\sigma')+ \nonumber \\
 &+&\partial_\sigma
 (r-s)^{AB}(\sigma,\Lambda,\Gamma)
 \delta(\sigma-\sigma')-2s^{AB}(\sigma,
 \Lambda,\Gamma)\partial_\sigma
 \delta(\sigma-\sigma') \ .  \nonumber \\
\end{eqnarray}
Using of the form of the Poisson
bracket (\ref{pbmA1a}) we can calculate
the algebra of monodromy matrices with
distinct  intervals
\begin{eqnarray}\label{algTT}
& &
\pb{\mT_{\alpha\beta}(\sigma,\sigma',\Lambda),
\mT_{\gamma\delta}(\xi,\xi',\Gamma)}=
\nonumber \\
&=&\mT_{\alpha\sigma_1}(\sigma,x_0,\Lambda)
\mT_{\gamma\sigma_2}(\xi,x_0,\Gamma)
\left[r(x_0,\Lambda,\Gamma)+\epsilon(\sigma-\xi)
s(x_0,\Lambda,\Gamma)\right]_{\sigma_1\sigma_2,\rho_1\rho_2}
\times \nonumber \\
&\times &
\mT_{\rho_1\beta}(x_0,\sigma',\Lambda)
\mT_{\rho_2,\delta}(x_0,\xi',\Gamma)-\nonumber \\
&-&\mT_{\alpha\sigma_1}(\sigma,y_0,\Lambda)
\mT_{\gamma\sigma_2}(\xi,y_0,\Gamma)
\left[r(y_0,\Lambda,\Gamma)
+\epsilon(\xi'-\sigma')s(y_0,\Lambda,\Gamma)
\right]_{\sigma_1\sigma_2,\rho_1\rho_2}
\times \nonumber \\
&\times&
\mT_{\rho_1\beta}(y_0,\sigma',\Lambda)
\mT_{\rho_2\delta}(y_0,\xi',\Gamma) \ ,
\nonumber \\
\end{eqnarray}
where $\epsilon(x)=\mathrm{sign}(x)$
and where we presume that $\sigma$ and
$\xi$ are larger than $\sigma'$ and
$\xi'$,
$x_0=\mathrm{min}(\sigma,\xi),\quad
y_0=\mathrm{max}(\sigma',\xi')$. It is
important to note that in the
non-ultralocal case the algebra
(\ref{algTT}), due to the presence of
the $s$-term, the function
\begin{equation}
\triangle^{(1)}(\sigma,\sigma',
\xi,\xi',\Lambda,\Gamma)=
\pb{\mT_{\alpha\beta}(\sigma,\sigma',\Lambda),
\mT_{\gamma\delta}(\xi,\xi',\Gamma)}
\end{equation}
is well defined and continuous where
$\sigma,\sigma',\xi,\xi'$ are all
distinct,but it has discontinuities
proportional to $2s$ across the
hyperplanes corresponding to some of
the $\sigma,\sigma',\xi,\xi'$ being
equal. Then if we want to define the
Poisson bracket of transfer matrices
for coinciding intervals ($\sigma=\xi,
\sigma'=\xi'$) or adjacent intervals
$(\sigma'=\xi \ \mathrm{or} \
\sigma=\xi')$ requires the value of the
discontinuous matrix-valued function
$\triangle^{(1)}$ at its
discontinuities.   It was shown in
\cite{Maillet:1985ec} that requiring
anti-symmetry of the Poisson bracket
and the derivation rule to hold imposes
the symmetric definition of
$\triangle^{(1)}$ at its discontinuous
points. For example, at $\sigma=\xi$ we
must define
\begin{eqnarray}
\triangle^{(1)}(\sigma,\sigma',\sigma,
\xi',\Lambda,\Gamma)=
\lim_{\epsilon\rightarrow 0^+}
\frac{1}{2} (\triangle^{(1)}
(\sigma,\xi,\sigma+
\epsilon,\xi',\Lambda,\Gamma)+
\triangle^{(1)}(\sigma,\sigma',
\sigma-\epsilon,\xi',\Lambda,\Gamma))
\nonumber \\
\end{eqnarray}
and likewise for all other possible
coinciding endpoints. This definition
of $\triangle^{(1)}$ at its
discontinuities implies an definition
of the Poisson bracket between
transition matrices for coinciding and
adjacent intervals that is consistent
with the anti-symmetry of the Poisson
bracket and the derivation rule.
However as was shown in
\cite{Maillet:1985ek} \footnote{For
very nice recent discussion, see
\cite{Dorey:2006mx}.} this definition
of the Poisson bracket
$\pb{\mT_{\alpha\beta}(\Lambda),
 \mT_{\gamma\delta}(\Gamma)}$
does not satisfy the Jacobi identity so
that in fact no strong definition of
the bracket
$\pb{\mT_{\alpha\beta}(\Lambda),
 \mT_{\gamma\delta}(\Gamma)}$
with coinciding or adjacent intervals
can be given without violating the
Jacobi identity. However, as was shown
in \cite{Maillet:1985ek} it is possible
to give a weak definition of this
bracket for coinciding or adjacent
intervals as well. We are not going
into details of the procedure,
interesting reader can read the
original paper \cite{Maillet:1985ek} or
more recent \cite{Dorey:2006mx}.  Let
us now define
\begin{equation}
\Omega_{\alpha\beta}(\Lambda)=\mT_{\alpha\beta}
(\infty,-\infty,\Lambda)
\end{equation}
Using the regularization procedure
developed in \cite{Maillet:1985ek} one
can then show that the Poisson bracket
between $\Omega_{\alpha\beta}(\Lambda)$
and $\Omega_{\gamma\delta}(\Gamma)$
takes the form
\begin{eqnarray}\label{pbOmega}
\pb{\Omega_{\alpha\beta}(\Lambda),
\Omega_{\gamma\delta}(\Gamma)}=
r_{\alpha\gamma,\sigma_1\sigma_2}(\Lambda,\Gamma)
\Omega_{\sigma_1\beta}(\Lambda)
\Omega_{\sigma_2\delta}(\Gamma)
-\Omega_{\alpha\sigma_1}(\Lambda)
\Omega_{\gamma\sigma_2}
(\Gamma)r_{\sigma_1\sigma_2,\beta\delta}(\Lambda,\Gamma)+\nonumber \\
+\Omega_{\alpha\sigma_1}(\Lambda)
s_{\sigma_1\gamma,\beta\sigma_2}(\Lambda,\Gamma)
\Omega_{\sigma_2\delta}(\Gamma)-\nonumber \\
-\Omega_{\gamma\sigma_1}(\Gamma)s_{\alpha\sigma_1,
\beta\sigma_2}(\Lambda,\Gamma)
\Omega_{\sigma_2\beta}(\Lambda) \ ,
\nonumber \\
\end{eqnarray}
where $r(\Lambda,\Gamma)\equiv
\lim_{\sigma\rightarrow \infty}
r(\Lambda,\Gamma,\sigma) \ ,
s(\Lambda,\Gamma)\equiv \lim_{\sigma
\rightarrow \infty}
s(\Lambda,\Gamma,\sigma)$. Using
(\ref{pbOmega}) we finally obtain
\begin{eqnarray}
\left\{\tr \Omega(\Lambda)\right.&,&
\left. \tr
\Omega(\Gamma)\right\}=
\pb{\Omega_{\alpha\alpha}(\Lambda),
\Omega_{\gamma\gamma}(\Gamma)}=\nonumber \\
&=&r_{\alpha\gamma,\sigma_1\sigma_2}(\Lambda,\Gamma)
\Omega_{\sigma_1\alpha}(\Lambda)
\Omega_{\sigma_2\gamma}(\Gamma)
-r_{\alpha\gamma,\sigma_1\sigma_2}(\Lambda,\Gamma)
\Omega_{\sigma_1\alpha}(\Lambda)
\Omega_{\sigma_2\gamma}(\Gamma)+
\nonumber \\
&+&\Omega_{\alpha\sigma_1}(\Lambda)
s_{\sigma_1\gamma,\alpha\sigma_2}(\Lambda,\Gamma)
\Omega_{\sigma_2\gamma}(\Gamma)
-\Omega_{\alpha\sigma_1}(\Lambda)
s_{\sigma_1\gamma,\alpha\sigma_2}(\Lambda,\Gamma)
\Omega_{\sigma_2\gamma}(\Gamma)=0 \ .
\nonumber \\
\end{eqnarray}
In other words we obtain that the
theory contains infinite number of
conserved charges that are in
involution. This fact implies a
classical integrability of given
theory.

\end{appendix}


\end{document}